\documentstyle[12pt]{article}
\def\beq{\begin{equation}}
\def\eeq{\end{equation}}
\def\bea{\begin{eqnarray}}
\def\eea{\end{eqnarray}}
\def\nn{\nonumber}
\def\ba{\begin{array}}
\def\ea{\end{array}}

\def\v{\vert}

\def\r{\rangle}

\begin{document}
% \begin{flushright} hep-th/9701097
% \end{flushright}
%\rightline{TIFR/TH/97-01}
%\rightline{January 1997}
\baselineskip16pt
\smallskip
\begin{center}
{\large \bf \sf
Bound and anti-bound soliton states for a quantum integrable \\
derivative nonlinear Schr\"odinger model }

\vspace{1.3cm}

{\sf B. Basu-Mallick$^1$\footnote{ 
e-mail address: biru@theory.saha.ernet.in },
Tanaya Bhattacharyya$^1$\footnote{e-mail address:
tanaya@theory.saha.ernet.in} and Diptiman Sen$^2$\footnote
{e-mail address: diptiman@cts.iisc.ernet.in } }

\bigskip

{\em $^1$Theory Group, \\
Saha Institute of Nuclear Physics, \\
1/AF Bidhan Nagar, Kolkata 700 064, India } 

\bigskip

{\em $^2$Centre for Theoretical Studies, Indian Institute of Science, \\
Bangalore 560012, India}
\end{center}

\bigskip

\vspace {1.1 cm}
\baselineskip=18pt
\noindent {\bf Abstract }

We find that localized quantum $N$-body soliton states exist for a derivative
nonlinear Schr\"odinger (DNLS) model within an extended range 
of coupling constant ($\xi_q$) given by $0 < \v \xi_q \v < \frac{1}{\hbar} 
\tan \left(\frac{\pi}{N-1}\right)$. We also observe that soliton states with 
both positive and negative momentum can appear for a fixed value of $\xi_q$.
Thus the chirality property of classical DNLS solitons is not preserved at 
the quantum level. Furthermore, it is found that the solitons with positive 
(negative) chirality have positive (negative) binding energy.

\medskip

\baselineskip=16pt
\vspace {.6 cm}
%\noindent PACS No. : 11.10.Lm, 11.30.-j, 02.30.Ik, 03.65.Fd 

\vspace {.1 cm}
%\noindent Keywords : Derivative nonlinear 
%Schr\"odinger model, Yang-Baxter equation, 
%Algebraic Bethe ansatz, Soliton 

\newpage

\baselineskip=18pt
\noindent \section {Introduction }
\renewcommand{\theequation}{1.{\arabic{equation}}}
\setcounter{equation}{0}

\medskip

Quantum soliton states associated with integrable field models
in 1+1 dimension have attracted a lot of interest in recent years [1-10]. 
These quantum soliton states are usually constructed by using the 
approach of either the coordinate Bethe ansatz or the algebraic Bethe 
ansatz. By applying the coordinate Bethe ansatz to an integrable
nonrelativistic Hamiltonian, one can directly obtain its exact eigenfunctions
in the coordinate representation. If such an eigenfunction 
decays sufficiently fast when a relative coordinate
measuring interparticle distance tends towards
infinity, we define this localized square-integrable eigenfunction 
as a quantum soliton state. It is also possible to construct 
quantum soliton states through the approach of the algebraic Bethe ansatz, 
by choosing appropriate distributions of the 
spectral parameter on the complex plane [2-4]. The stability of quantum soliton
states, in the presence of small external perturbations, can be 
measured by calculating their binding energy. 
It is usually found that localized quantum soliton states of 
various integrable models, including the well known nonlinear 
Schr\"odinger model and sine-Gordon model, 
have positive binding energy [1-5]. 

The purpose of the present article is to study the quantum soliton 
states of an integrable derivative nonlinear Schr\"odinger 
(DNLS) model [6-10]. Classical and quantum versions of 
DNLS model have found applications in different areas of
physics like circularly polarized nonlinear Alfven waves in a plasma [11,12],
quantum properties of the optical solitons in fibers [13] and 
some chiral Luttinger liquids which are obtained by dimensional
reduction of a Chern-Simons model defined in two dimensions [14-15]. 
It is well known that the classical DNLS model can have 
solitons with momenta in only one direction [16-18]. Here we want to 
investigate whether this chirality property of the classical solitons 
is preserved at the quantum level. The Hamiltonian of quantum DNLS model
in its second quantized form is given by [6,7]
\beq
H= \hbar\int_{-\infty}^{+\infty} dx ~\Big[\, \psi_x^{\dagger}(x)\psi_x(x) +
i\xi_q \{{\psi^\dagger}^2(x)\psi(x)\psi_x(x) - 
\psi^\dagger_x(x) {\psi^\dagger}(x) \psi^2(x) \} \,\Big] \, ,
\label {a1}
\eeq
where we have set the particle mass $m = 1/2$, $\xi_q$ is the coupling 
constant, and $ \psi(x),~ \psi^\dagger(x)$ are 
field operators obeying the following equal time commutation relations:
$ [ \psi(x), \psi(y) ] = [ \psi^\dagger(x), \psi^\dagger(y) ] = 0,$ 
$ [ \psi(x), \psi^\dagger(y) ] $ $ = \hbar \delta (x-y). $ Quantum soliton 
states of this DNLS model have been constructed through the approach of the
algebraic Bethe ansatz [8-10] as well as the coordinate Bethe ansatz [6,7].
By applying the coordinate Bethe ansatz, it is found that quantum $N$-body
soliton (henceforth called $N$-soliton) states exist for this DNLS model 
provided that $\xi_q$ is restricted within the range: 
$0 < \v \xi_q \v < \frac{1}{\hbar} \tan \left(\frac{\pi}{N}\right)$.
Moreover it is observed that, similar to the classical case,
such $N$-soliton states can have only positive momentum for the case 
$\xi_q >0$ [6]. But, it was mentioned later that, soliton states 
can exist even for negative momentum for small values of $N$ [7]. 
However, no explicit range of $\xi_q $ is known for which such $N$-solitons 
with negative momenta may appear in the quantum DNLS model.

In this article we reinvestigate
the range of $\xi_q$ for which localized quantum $N$-soliton
states exist in the case of DNLS model. In Sec. 2, we apply the
coordinate Bethe ansatz for this purpose and show that quantum $N$-soliton
states exist within an extended range given by $0 < \v \xi_q \v < 
\frac{1}{\hbar} \tan \left(\frac{\pi}{N-1}\right)$. We also find that the 
momentum of the $N$-soliton state has the same sign as the coupling constant 
when it lies within the range $0 < \v \xi_q \v< \frac{1}{\hbar}\tan 
\left(\frac{\pi}{N}\right)$ and has the opposite sign to the coupling constant
when $ \frac{1}{\hbar}\tan \left(\frac{\pi}{N}\right) 
<\v \xi_q \v< \frac{1}{\hbar}\tan \left(\frac{\pi}{N-1}\right)$. As a result,
soliton states with both positive and negative momentum can appear for 
a fixed value of $\xi_q$. In Sec. 3 we find out the binding energy of 
these soliton states. Sec. 4 is the concluding section. 

\vspace{1cm}

\noindent \section {Quantum N-soliton states for DNLS model }
\renewcommand{\theequation}{2.{\arabic{equation}}}
\setcounter{equation}{0}

\medskip

For applying the coordinate Bethe ansatz, we separate the full bosonic Fock
space associated with Hamiltonian (\ref {a1})
into disjoint N-particle subspaces $\vert F_N \rangle $.
We are interested in solving the eigenvalue equation $H \v F_N \r = E \v F_N 
\r $. The coordinate representation of this equation yields
\bea
H_N \, \tau_N( x_1, x_2, \cdots , x_N ) \, = \, 
E \, \tau_N( x_1, x_2, \cdots , x_N ) \, ,
\label{b1}
\eea
where the N-particle symmetric wave function $ \tau_N( x_1, x_2, \cdots , x_N)$
is defined as 
\bea
\tau_N( x_1, x_2, \cdots , x_N ) = 1/\sqrt{n!}\langle 0 \vert \psi(x_1)\cdots
\psi(x_N)\vert F_N \rangle \, ,
\label{b2}
\eea 
and $H_N$, the projection of the second-quantized Hamiltonian $H$ (\ref {a1}) 
on the N particle sector, is given by
\beq
H_N = -\hbar^2 ~\sum_{j=1}^N\frac{\partial^2}{\partial x_j^2} + 2i \hbar^3 
\xi_q ~\sum_{l<m}\delta(x_l - x_m)\Big( \frac{\partial}{\partial x_l} 
+ \frac{\partial} {\partial x_m} \Big).
\label {b3}
\eeq
It is evident that this $H_N$ commutes with 
the total momentum operator in the $N$-particle sector, which is defined as
\beq
P_N= -i\hbar\sum_{j=1}^N \frac{\partial}{\partial x_j} \, . 
\label {b4}
\eeq
Next, we divide the coordinate space $R^N \equiv \{ x_1, x_2, \cdots x_N \} $ 
into various $N$-dimensional sectors defined through inequalities like 
$x_{\omega(1)}< x_{\omega(2)}< \cdots < x_{\omega(N)}$, where $\omega(1), 
\omega(2), $ $\cdots , \omega(N)$ represents a permutation of $1,2, \cdots ,N$.
Within each such sector the interaction potential of the Hamiltonian 
(\ref {b3}) is zero and the resulting eigenfunction is just a superposition
of the free particle wave functions. Coefficients associated with 
these free particle wave functions can be obtained through the 
interaction part of the Hamiltonian (\ref {b3}), which is nontrivial 
at the boundary of two adjacent sectors. It is known that all such necessary 
coefficients for the Bethe ansatz solution of a $N$-particle 
system can be obtained by solving the corresponding two particle
problem [19]. Hence let us first try to construct the eigenfunctions 
of the Hamiltonian (\ref {b3}) for the two particle case, without 
imposing any symmetry property on $\tau_2(x_1,x_2)$ under the exchange 
of particle coordinates. For 
$x_1<x_2 $ region, we may take such eigenfunction as
\bea
\tau_2(x_1, x_2) = \exp \, \{ i(k_1 x_1 + k_2 x_2) \} \,
\label{b5}
\eea
where $k_1$ and $k_2$ are two distinct wave numbers. 
Using eqn. (\ref {b1}) for $N=2$, one can show that this two-particle 
wave function will take the following form in the region $x_1>x_2 \, $:
\bea
\tau_2(x_1,x_2) = 
A(k_1,k_2)\exp \, \{ i(k_1 x_1 + k_2 x_2) \} + B(k_1,k_2)
\exp \{ i(k_2 x_1 + k_1 x_2) \} \, ,
\label{b6}
\eea
where the `matching coefficients' $A(k_1,k_2)$ and $B(k_1,k_2)$ are given by
\beq
A(k_1,k_2)=
\frac{(k_1-k_2)+i\hbar\xi_q(k_1+ k_2)}{(k_1-k_2)} \, , ~~~~
B(k_1,k_2) = 1- A(k_1 , k_2) \, . 
\label{b7}
\eeq
By using these matching coefficients, one can construct completely symmetric
N-particle eigenfunctions for the Hamiltonian (\ref {b3}). In the region
$x_1< x_2 < \cdots < x_N$, such eigenfunctions are given by [6,19]
\beq
\tau_N(x_1, x_2 , \cdots , x_N)= \sum_\omega\left (\prod
_{l<m}\frac{A(k_{\omega(m)},k_{\omega(l)})}{A(k_m,k_l)}\right)
\rho_{\omega(1), \omega(2), \cdots , \omega(N)}
(x_1, x_2, \cdots , x_N) \, ,
\label{b8}
\eeq
where 
\beq
\rho_{\omega(1), \omega(2), \cdots , \omega(N)}
(x_1, x_2, \cdots , x_N) =
\exp \, \{ i (k_{\omega(1)}x_1 + \cdots + k_{\omega(N)} x_N ) \} \, .
\label{b9}
\eeq
In the expression (\ref {b8}), $k_n$s are all distinct wave numbers 
and $\sum_{\omega}$ implies summing over all
permutations of the numbers (1,2,....N). The eigenvalues of 
the momentum (\ref {b4}) and Hamiltonian (\ref {b3}) operators, corresponding 
to the eigenfunctions $\tau_N(x_1, x_2, \cdots , x_N)$, are given by
\bea
&&~~~~~~~~P_N \, \tau_N(x_1, x_2, \cdots , x_N) = \hbar 
\Big(\sum_{j=1}^N k_j \Big) \, \tau_N(x_1, x_2, \cdots , x_N) \, , \nn \\
&&~~~~~~~~H_N \, \tau_N(x_1, x_2, \cdots , x_N) = 
\hbar^2 \Big(\sum_{j=1}^N k_j^2 \Big) \,
\tau_N(x_1, x_2, \cdots , x_N) \, . \nn ~~~~~~~~~~~~~~~~~(2.10a,b)
\eea
\addtocounter{equation}{1}

The wave function (\ref{b8})
will represent a localized soliton state, if it decays 
when any of the relative coordinates measuring an interparticle distance
tends towards infinity. To obtain the condition for the existence of 
such localized soliton state, let us consider the following
wave function in the region $x_1<x_2<\cdots <x_N$ :
\bea
\rho_{1,2, \cdots ,N } (\, x_1,x_2, \cdots , x_N \,)
= \exp \,(i\sum_{j=1}^N k_j x_j) .
\label{b11}
\eea
As before, the momentum eigenvalue corresponding to this 
wave function is given by $\hbar\sum_{j=1}^N k_j $.
Since this momentum eigenvalue must be a real quantity, one gets the condition
\beq
\sum_{j=1}^N q_j = 0 \, ,
\label{b12}
\eeq
where $q_j$ denotes the imaginary part of $k_j$. 
The probability density corresponding to the wave function $\rho_{1,2, 
\cdots ,N} (\, x_1,x_2, \cdots , x_N \,)$ (\ref{b11}) can be expressed as 
\bea
{|\rho_{1,2,\cdots ,N} (\, x_1,x_2, \cdots , x_N \,)|}^2 
= \exp \, \Big\{ 2 \sum_{r=1}^{N-1} \Big(\sum_{j=1}^r q_j\Big)y_r 
\Big \} \, ,
\label{b13}
\eea
where the $y_r$'s are the $N-1$ number of relative coordinates: $y_r \equiv 
x_{r+1} - x_r$. It is evident that this probability density (\ref{b13}) 
decays exponentially in the limit $y_r \rightarrow \infty $ provided that 
all the following conditions are satisfied:
\beq
q_1< 0 \, , ~~~~q_1+q_2 < 0 \, , ~~~\cdots\cdots ~~~ ,
\sum_{j=1}^{N-1} q_j < 0 \, . 
\label{b14}
\eeq
Note that the wave function (\ref {b11}) is obtained by taking $\omega$
as the identity permutation in (\ref{b9}). However, the
full wave function (\ref {b8}) also contains terms like (\ref {b9}) with 
$\omega$ representing all possible nontrivial permutations. 
The conditions which ensure the decay of such a term 
with nontrivial permutation, in general, contradict the 
conditions (\ref {b14}). Consequently, to have an
overall decaying wave function (\ref {b8}), the coefficients of all terms 
$\rho_{\omega(1), \omega(2), \cdots , \omega(N)}
(x_1, x_2, \cdots , x_N)$ with nontrivial permutations 
are required to vanish. This requirement leads to a set of relations like
\beq
A( k_{1}, k_{2} ) = 0, ~~A( k_{2}, k_{3} ) = 0, ~\cdots\cdots ~, A(
k_{N-1}, k_{N}) = 0 .
\label{b15}
\eeq
Thus the simultaneous validity of the conditions
(\ref{b12}), (\ref{b14}) and (\ref{b15}) ensures that the full wave function
$\tau_N(x_1, x_2, \cdots , x_N)$ (\ref {b8}) would represent a localized
soliton state. Using the conditions (\ref {b12}) and (\ref{b15}), one can
derive the expression for the complex $k_n$'s as
\beq
k_n = \chi \, e^{-i(N+1-2n)\phi} \, ,
\label{b16}
\eeq
where $\chi$ is an yet undetermined real parameter, and $\phi$ is another
real parameter which is related to the coupling constant as 
\beq
\phi=\tan^{-1}\left( \hbar\xi_q \right) \, .
\label{b17}
\eeq
To obtain a unique value of $\phi$ from the above equation, we restrict it
within the fundamental region: $-\frac{\pi}{2} < \phi < \frac{\pi}{2} $. 
In this context, it may be mentioned that a relation equivalent to (\ref{b16})
can also be obtained through the method of the algebraic Bethe ansatz,
when quantum soliton states of DNLS model are considered [8-10]. Next, let us 
verify whether these $k_n$s (\ref {b16}) satisfy the conditions (\ref{b14}). 
Summing over the imaginary parts of these $k_n$s, we can express 
the conditions (\ref {b14}) in the form 
\beq
\chi \frac{\sin \phi l}{\sin \phi} \sin\{\phi(N-l)\} > 0 \, ,
\label{b18}
\eeq
where $l \in [1,2,3, \cdots , N-1 ]$. For the simplest case $N=2$, 
the above condition is satisfied when $\phi $ lies within the range
$0 < \phi < \frac {\pi}{2}$ ($- \frac {\pi}{2} < \phi < 0$)
for the choice $\chi >0$ ($\chi < 0$). Thus any value of $\phi$
within its fundamental region can generate a $2$-soliton state. 
Next, we consider the case where $N \geq 3$.
It may be noted that, $\sin \phi$, $\sin \phi l$, and $\sin\{\phi( N-l)\}$ 
are all positive (negative) when $\phi$ takes value 
within the range $0<\phi< \frac{\pi}{N-1}$ 
($-\frac{\pi}{N-1}<\phi<0$). Hence the condition (\ref {b18}) will be 
satisfied for the choice $\chi >0$ ($\chi < 0$),
when $\phi$ lies within the range $0<\phi< \frac{\pi}{N-1}$ ($-
\frac{\pi}{N-1}<\phi<0$). Due to eqn. (\ref {b17}),
the corresponding coupling constant $\xi_q$ lies within
the range $0<\xi_q<\frac{1}{\hbar}\tan\frac{\pi}{N-1}$
(\,$- \frac{1}{\hbar}\tan\frac{\pi}{N-1}<\xi_q<0 $ \,).
As a result, we get localized $N$-soliton states within the above 
mentioned range of $\phi$ or $\xi_q$.
Earlier it has been shown that $N$-soliton state can be obtained 
for the case $\xi_q >0$ when $\phi$ lies within the range
$0<\phi< \frac{\pi}{N}$ [6]. The present analysis yields a wider range of
$\phi$ for which a $N$-soliton state can be constructed.

Next, by using eqns. (\ref {b16}) and (2.10a), we obtain the momentum 
eigenvalue ($P$) corresponding to these soliton states as
\beq
P = \hbar\chi \,\frac{\sin \phi N}{\sin \phi} \, .
\label{b19}
\eeq
Thus, for any given value of $\phi$, the momentum of such $N$-soliton state 
is proportional to the value of parameter $\chi$. 
Due to eqn. (\ref {b19}), it is evident that $P$ and $\chi$ have the same 
sign when $\phi$ lies in the range 
$0 < \v \phi \v < \frac{\pi}{N}$ and the opposite sign when $\phi$ lies in the
range $\frac{\pi}{N} < \v \phi \v < \frac {\pi}{N-1}$. However, we have 
seen earlier that $\chi$ and $\xi_q$ must
have the same sign within the allowed range of a
$N$-soliton state. Consequently, the momentum of the
$N$-soliton state has the same sign as $\xi_q$ when 
$0 < \v \phi \v< \frac{\pi}{N}$ and has the opposite sign to $\xi_q$
when $\frac{\pi}{N} < \v \phi \v < \frac {\pi}{N-1}$. We may call solitons
with momenta having the same (opposite) sign as 
$\xi_q$ as solitons with positive (negative) chirality. 
It may be noted that the range of $\phi$ given by 
$\frac{\pi}{N} < \v \phi \v < \frac {\pi}{N-1}$ 
does not lie within its fundamental region for the case $N=2$. As a result, 
$N$-soliton states with negative chirality can appear only for $N \geq 3$. 

The appearance of both positive and 
negative momentum solitons in quantum DNLS model can be described most
conveniently by considering a fixed positive value of $\phi$ or $\xi_q$.
Let us first take a fixed value of $\phi$ so that $\frac{\pi}{\phi}$ is an 
integer. In this case we obtain solitons with positive momentum 
for the choice $N= 2,3,
\cdots, \frac{\pi}{\phi} -1 $ and a single soliton with zero momentum
for $N_{max}= \frac{\pi}{\phi}$. Next we consider the
case where $\frac{\pi}{\phi}$ is non-integer and 
denote its integral part by $\left[ \frac{\pi}{\phi} \right]$.
In this case we obtain solitons with positive momentum 
for the choice $N= 2,3, \cdots, \left [ \frac{\pi}{\phi} \right] $ 
and a single soliton with negative
momentum for $N_{max}= \left [\frac{\pi}{\phi} \right]+1 $. 
Thus we find that the chirality property of classical DNLS solitons
is broken at the quantum level.
It may be further noted that the ratio of solitons with
positive and negative momentum is of the order of $\frac{\pi}{\phi}$.
Moreover, due to eqn. (\ref {b17}), we get $\phi \rightarrow 0$ at the 
classical $\hbar \rightarrow 0$ limit (for any fixed value of $\xi_q$).
As a result, the ratio of solitons with positive and negative momentum 
becomes infinitely large at the classical limit.

\vspace{1cm}

\noindent \section {Binding energy of a $N$-soliton state}
\renewcommand{\theequation}{3.{\arabic{equation}}}
\setcounter{equation}{0}

\medskip

Here we want to calculate the binding energy for the 
$N$-soliton states of quantum DNLS model within the allowed 
range of $\phi$ given by $0 < \v \phi \v < \frac{\pi}{N-1}$. 
Substituting the values of complex $k_n$ (\ref {b16}) in (2.10b), we obtain 
the energy eigenvalue ($E$) corresponding to these $N$-soliton states as
\bea
E =\frac{\hbar^2 \chi^2\sin(2\phi N)}{\sin(2\phi)} \, .
\label{c1}
\eea
To calculate binding energy, we assume that the momentum $P$ 
(\ref {b19}) of the $N$-soliton state is equally distributed among $N$ number 
of single-particle scattering states. The real 
wave number associated with each of these single-particle states is denoted 
by $k_0$.  With the help of eqns. (2.10a) and (\ref{b19}), we obtain 
\beq
k_0=\frac{\chi\sin(\phi N)}{N\sin \phi}. 
\label{c2}
\eeq
Using eqn. (2.10b), one can easily calculate the total energy for 
$N$ number of such single-particle scattering states as
\bea
E'= \hbar^2 N k_0^2 = \frac{\hbar^2 \chi^2\sin^2\phi N}{N\sin^2 \phi} \, .
\label{c3}
\eea
Subtracting $E$ (\ref {c1}) from $E'$ (\ref {c3}), we obtain the 
binding energy of quantum $N$-soliton state as 
\beq
E_B (\phi, N) \equiv E'- E 
= \frac{\hbar^2 \chi^2\sin \phi N}{\sin \phi}\Big\{\frac{
\sin \phi N}{N\sin \phi} - \frac{\cos \phi N}{\cos \phi} \Big\} .
\label{c4}
\eeq
It may be noted that the above expression of binding energy remains 
invariant under the transformation $\phi \rightarrow -\phi$. Thus it 
is sufficient to investigate the nature of this binding energy only for
positive values of $\phi$ within the range $0< \phi < \frac{\pi}{N-1}$. 
Substituting $N=2$ to eqn. (\ref {c4}),
we obtain $E_B(\phi, 2) =2\hbar^2 \chi^2\sin^2 \phi$. Thus we get
$E_B (\phi,2) >0$ for any nonzero value of $\phi$. 
For N=3, eqn. (\ref{c4}) takes the form
$E_B(\phi, 3) = \frac{8\hbar^2 \chi^2}{3}\sin^2\phi(3-4\sin^2\phi)$.
Here we get $E_B (\phi,3)>0$ when $0< \phi <\frac{\pi}{3}$ and 
$E_B(\phi, 3)<0$ when $\frac{\pi}{3}< \phi < \frac{\pi}{2} $.
So from the above examples we can guess that $E_B(\phi,N)>0$ when
$0<\phi<\frac{\pi}{N}$ and $E_B<0$ when $\frac{\pi}{N}< \phi < 
\frac{\pi}{N-1}$. By using the method of induction, 
we will try to prove that this condition is really valid 
 for arbitrary values of $N$. To this end, we write 
$E_B (\phi, N) $ (\ref {c4}) in the form 
\beq
E_B (\phi, N) = \frac{\hbar^2 \chi^2\sin \phi N}{\sin \phi} f(\phi, N) \, ,
\label {c5}
\eeq
where
\beq
f(\phi, N) = \frac{
\sin \phi N}{N\sin \phi} - \frac{\cos \phi N}{\cos \phi} \, .
\label{c6}
\eeq
To use the method of induction, let us first assume that
$f(\phi, N) > 0 $, i.e.
\beq
\frac{\sin \phi N}{\sin \phi} > \frac{N\cos \phi N}{\cos \phi} \, ,
\label {c7}
\eeq
when $\phi$ lies within the range $0<\phi<\frac{\pi}{N}$. 
%We shall prove that
%$f(\phi, N+1)>0$ for $0<\phi<\frac{\pi}{N}$. 
Next we consider an identity given by 
\bea
\frac{\sin\phi(N+1)}{\sin\phi}
= \left(\frac{\sin \phi N}{\sin \phi}\right)\cos \phi + \cos \phi N . \nn
\eea
Using the inequality (\ref {c7}) and the above identity, we obtain 
\bea
\frac{\sin\phi(N+1)}{\sin \phi}>(N+1)\cos\phi N \, ,
\label {c8}
\eea
for $0<\phi<\frac{\pi}{N}$. Next we consider another identity given by
\bea
\cos\phi N = \frac{\cos\phi (N+1)}{\cos\phi}+
\frac{\sin \phi\sin(\phi N)}{\cos \phi}. \nn
\eea
Since ${\sin\phi\sin(\phi N)}/{\cos\phi}>0$ for $0<\phi
<\frac{\pi}{N}$, the above identity yields
\bea
(N+1)\cos\phi N >\frac{(N+1)\cos\phi (N+1)}{\cos \phi} \, ,
\label{c9}
\eea
for $0<\phi<\frac{\pi}{N}$.
Now, combining the inequalities (\ref {c8}) and (\ref {c9}), we get
\bea
\frac{\sin\phi(N+1)}{\sin\phi}>\frac{(N+1)\cos\phi (N+1)}{\cos \phi} \, ,
\label{c10}
\eea
for $0<\phi<\frac{\pi}{N}$. Thus by assuming that $f(\phi, N) > 0 $
within the range $0<\phi<\frac{\pi}{N}$, we are able to show that 
$f(\phi, N+1) > 0 $ within the same range of $\phi$. By putting $N=2$ 
in eqn. (\ref {c6}), it is easy to check that $f(\phi,2)>0$ for
$0<\phi<\frac{\pi}{2}$.  Thus by using the method of induction, it is 
established that $ f(\phi, N)>0$ within the range 
$0<\phi<\frac{\pi}{N-1}$ for all $N\geq 3$. Now, it may be noted that
$\frac{\sin\phi N}{\sin \phi} > 0$ for $0<\phi<\frac{\pi}{N}$.
Consequently, from eqn. (\ref{c5}) we find that the binding energy 
for a $N$-soliton state of quantum DNLS model is positive when $\phi$ 
lies in the range $0<\phi<\frac{\pi}{N}$. However, it may also be noted that 
$\frac{\sin\phi N}{\sin \phi}<0$ within the
range $\frac{\pi}{N}<\phi<\frac{\pi}{N-1}$.
As a result, from eqn. (\ref{c5}) we interestingly find that the binding 
energy for a $N$-soliton state is negative when $\phi$ lies 
in the range $\frac{\pi}{N}<\phi <\frac{\pi}{N-1}$. Thus we find 
examples of quantum solitons in an integrable model where 
the binding energy can be negative. 

\medskip

\noindent \section {Concluding Remarks}

By applying the coordinate Bethe ansatz, we have investigated
the range of the coupling constant $\xi_q$ for which localized
quantum $N$-soliton states exist for DNLS model. It turns out that such 
quantum $N$-soliton states exist within an extended range of $\xi_q$ given by
$0 < \v \xi_q \v < \frac{1}{\hbar} \tan \left(\frac{\pi}{N-1}\right)$.
We also observe that the $N$-soliton state has positive chirality when 
$\xi_q$ lies within the range
$0 < \v \xi_q \v< \frac{1}{\hbar}\tan \left(\frac{\pi}{N}\right)$ and negative
chirality when $ \frac{1}{\hbar}\tan \left(\frac{\pi}{N}\right) 
<\v \xi_q \v< \frac{1}{\hbar}\tan \left(\frac{\pi}{N-1}\right)$. 
Thus we find that the chirality property of classical DNLS solitons is not 
preserved at the quantum level. We also calculate the binding energy for the 
quantum soliton states of this model. Interestingly, it is found 
that the binding energy for a $N$-soliton state is 
positive when $0 < \v \xi_q \v <\frac{1}{\hbar}\tan\left(\frac{\pi}{N}\right)$
and negative when $\frac{1}{\hbar}\tan \left(\frac{\pi}{N}\right)< \v 
\xi_q \v < \frac{1}{\hbar}\tan\left(\frac{\pi}{N-1}\right)$. 
Thus we can conclude that the solitons with positive chirality form bound 
states with positive binding energy. On the other hand, solitons with 
negative chirality have negative binding energy and hence they form anti-bound
states. Consequently, solitons with negative chirality
would be unstable in the presence of external perturbations. As a 
future study, it might be interesting to calculate the decay rate of such 
quantum solitons by introducing small perturbations to the DNLS Hamiltonian.

Recently, we have discovered additional ranges of values of $\phi$ for which
there are bound and anti-bound $N$-soliton states. The complete description
of these ranges can be found using concepts from number theory such as Farey
sequences and continued fractions [20].

\bigskip

\bigskip

%\noindent {\bf Acknowledgments }

% \newpage

\leftline {\large \bf References}
\medskip
\begin{enumerate}

\item H. B. Thacker, Rev. Mod. Phys. 53 (1981) 253.

\item L. D. Faddeev, Sov. Sci. Rev. C 1 (1980) 107; in {\it Recent Advances in
Field Theory and Statistical Mechanics,} ed. J.-B. Zuber and R.Stora,
(North-Holland, Amsterdam, 1984 ) p.561. 

\item E. K. Skylanin, in 
Yang-Baxter Equation in Integrable systems, Advanced series in Math. Phys. 
Vol. 10, edited by M. Jimbo (World Scientific, Singapore, 1990) p.121.

\item E. K. Skylanin, L. A. Takhtajan and L. D. Faddeev, Theor. Math. Phys.
40 (1980) 688.

\item G. Bhattacharya and S. Ghosh, Int. J. Mod. Phys. A 4 (1989) 627. 

\item A. G. Shnirman, B. A. Malomed and E. Ben-Jacob, Phys. Rev. A 50 (1994) 
3453.

\item D. Sen, {\it Quantization of the derivative nonlinear 
Schr\"odinger equation}, \hfil \break cond-mat/9612077.

\item A. Kundu and B. Basu-Mallick, J. Math. Phys. 34 (1993) 1052.

\item B. Basu-Mallick and T. Bhattacharyya, Nucl. Phys. B 634 [FS] (2002) 611.

\item B. Basu-Mallick and T. Bhattacharyya, {\it Jost solutions and 
quantum conserved quantities of an integrable derivative 
nonlinear Schr\"odinger model}, hep-th/0304063.

\item M. Wadati, H. Sanuki, K. Konno and Y.-H. Ichikawa, 
Rocky Mountain J. Math. 8 (1978) 323; Y.-H. Ichikawa and 
S. Watanabe, J. de Physique 38 (1977) C 6-15.

\item P. A. Clarkson, Nonlinearity 5 (1992) 453.

\item M. Rosenbluh and R. M. Shelby, Phys. Rev. Lett. 66 (1991) 153. 

\item U. Aglietti, L. Griguolo, R. Jackiw, S.-Y. Pi and D. Seminara, Phys.
Rev. Lett. 77 (1996) 4406; R. Jackiw, {\it A nonrelativistic chiral soliton
in one dimension}, hep-th/9611185.

\item S. J. Benetton Rabello, Phys. Rev. Lett. 76 (1996) 4007; (E) 77
(1996) 4851.

\item H.H. Chen, Y.C. Lee and C.S. Liu, Phys. Scr. 20 (1979) 490.

\item D.J. Kaup and A.C. Newell, J. Math. Phys. 19 (1978) 798.

\item H. Min and Q.-H. Park, Phys. Lett. B 388 (1996) 621.

\item E. Gutkin, Ann. Phys. 176 (1987) 22.

\item B. Basu-Mallick, T. Bhattacharyya and D. Sen, under preparation.

\end{enumerate}

\end{document}